\documentclass[runningheads]{llncs}
\usepackage{graphicx}
\usepackage{amsmath,amssymb} 
\usepackage{color}
\usepackage{float}
\usepackage{subfigure}
\usepackage{url}

\begin{document}

\newcommand{\point}{
    \raise0.7ex\hbox{.}
    }

\pagestyle{headings}

\mainmatter

%===========================================================
\title{Combined hapto-visual and auditory rendering of cultural heritage objects} % Replace with your title

\titlerunning{Combined hapto-visual and auditory rendering of cultural heritage objects} % Replace with your title

\authorrunning{P.~K.~Aniyath, K.~G.~Sreeni, P.~Kumari, S.~Chaudhuri} % Replace with your names

\author{Praseedha Krishnan Aniyath, Sreeni Kamalalayam Gopalan, Priyadarshini Kumari, Subhasis Chaudhuri} % Replace with your names
\institute{Vision and Image Processing Lab, Department of Electrical Engineering, IIT Bombay, India} % Replace with your institute's address

\maketitle
%===========================================================
\begin{abstract}
In this work, we develop a multi-modal rendering framework comprising of hapto-visual and auditory data. The prime focus is to haptically render point cloud data representing virtual 3-D models of cultural significance and also to handle their affine transformations. Cultural heritage objects could potentially be very large and one may be required to render the object at various scales of details. Further, surface effects such as texture and friction are incorporated in order to provide a realistic haptic perception to the users. Moreover, the proposed framework includes an appropriate sound synthesis to bring out the acoustic properties of the object. It also includes a graphical user interface with varied options such as choosing the desired orientation of 3-D objects and selecting the desired level of spatial resolution adaptively at runtime. A fast, point proxy-based haptic rendering technique is proposed with proxy update loop running $100$ times faster than the required haptic update frequency of $1$ kHz. The surface properties are integrated in the system by applying a bilateral filter on the depth data of the virtual 3-D models. Position dependent sound synthesis is incorporated with the incorporation of appropriate audio clips.
\end{abstract}

%===========================================================
\section{Introduction}\label{intro}
We interact with our physical world mostly through visual, auditory and tactile sensations. Haptics is an emerging research field which tries to emulate the latter sensation in the virtual world. Incorporation of haptics in the virtual environment provides a better immersive experience to the users, especially to the visually impaired persons. Further, a combined hapto-visual rendering enhances the realism of the haptic interaction even for the sighted users. Many authors have also suggested the inclusion of surface properties like texture and friction in the haptic domain for realistic haptic perception. However, in most cases of prior art, the rendering of surface properties such as texture and friction is not realistic enough to provide sufficient immersion to the users. This is more so as common haptic interface devices like Novint Falcon and SensAble Phantom are kinaesthetic and not tactile.

Haptic rendering methods like god object rendering algorithm ~\cite{Sali04}  work well in the case of polygon based representation of 3-D models.
On the other hand, it fails to properly render 3-D objects represented using point cloud data. Moreover, 3-D objects may appear at various different scales and orientations, and the user needs to experience the object at different levels of details. A polygon-based rendering scheme is not suitable to implement variations in scale and orientation. This is because recomputing the mesh structure during the interaction is not feasible as it is time consuming. The authors in~\cite{Sreeni12} did speed up the multi-level hapto-visual rendering using a Monge surface. However, their focus is mainly on fast rendering of a single valued function at different levels of details and cannot handle any rotation of the space and does not provide any surface friction.
\begin{figure}
\centering
\includegraphics[width=0.6\textwidth]{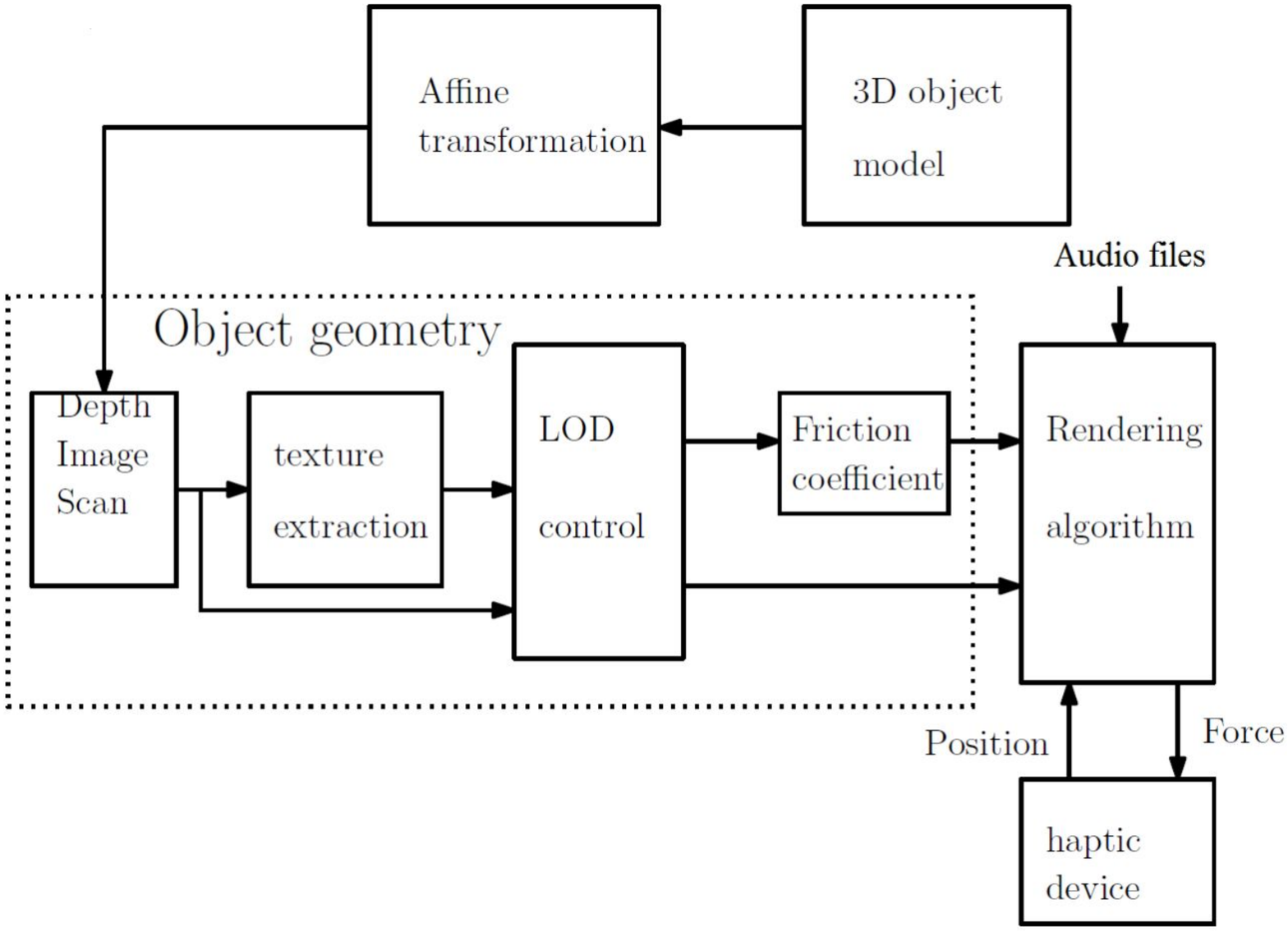}
\centering
\caption{Block diagram illustrating the proposed system. LoD stands for level of details.}
\label{block_fig}
\end{figure} 
This paper aims at enhancing the virtual immersion during haptic interaction with point cloud 3-D models by incorporating surface properties such as texture and friction. The coefficient of friction is computed from the texture component extracted from the scanned depth data. But for these properties, the virtual environment would often feel slippery since the direction of the force vector is always perpendicular to the surface and thus would not be sufficiently realistic. The new rendering framework can also handle affine transformations such as rotation, translation, expansion and geometric contraction of 3-D objects as illustrated in Fig.~\ref{block_fig}. Initially, we generate depth data at each point of the model at any desired orientation by reading the contents of depthbuffer in OpenGL using the inbuilt GLUT command \texttt{glReadPixels()} and create a Monge surface from it. This surface is then hapto-visually rendered at different scales adaptively at run time. We propose the use of a bilateral filter to extract the local surface texture and compute the dynamic friction as a function of the local texture. We show that the user's experience can be enriched by allowing the user to interact with the object at multiple resolutions. Audio rendering is also incorporated in the proposed system for enhanced virtual experience by playing appropriate audio files based on the position of the haptic probe inside the virtual environment. We also implement a graphical user interface for easier accessibility.

The organization of the paper is as follows. Section \ref{lit_sec} provides a review of the related literature. Section \ref{meth_sec} explains the proposed method. Haptic and graphic rendering techniques and additional functionalities are discussed in section \ref{render}. Section \ref{results} illustrates the results. The conclusions are summarized in section \ref{conclusions}.
\section{Literature Review}\label{lit_sec}
The basic haptic rendering technique is polygon based in which virtual objects are represented using polygonal meshes. In polygon based rendering, each time the haptic interface point (HIP) penetrates the object, the rendering algorithm  finds the closest point on the mesh defined surface and computes the penetration depth of HIP inside the object. If $\mathbf{x}$ is the vector representing the penetration depth, the reaction force is calculated as $\mathbf{F}=-k\mathbf{x}$, where $k$ is the stiffness constant of the surface. This method has problems in determining the appropriate force direction while rendering thin objects. Authors in~\cite{Sali95},~\cite{Rusp97} independently proposed the concept of god-object algorithm and proxy algorithm, respectively, to solve this problem. The god-object rendering algorithm includes another point in addition to the HIP, called ``god-object or proxy". In free space the proxy and the HIP are collocated. However, as the HIP penetrates the virtual object, the proxy is constrained to lie on the surface of the virtual object~\cite{Lay07}. But in the case of virtual objects represented using point cloud data, the proxy would slip into the object. It is called the ``fall-through" problem of the god-object. This can be avoided by using a spherical proxy instead of a point. However, an increase in radius of the proxy impairs the haptic interaction by smoothing out the surface texture. 

Many authors have also developed algorithms to render point cloud data directly. Lee \emph{et~al.} have proposed a rendering method with point cloud data which estimates the distance from HIP to the closest point on the moving least square surface defined by the given point set~\cite{Lee07}. El-Far~ \emph{et~al.} used axis aligned bounding boxes to fill the voids in the point cloud and then rendered with a god object rendering technique~\cite{Naim08}. Leeper~\emph{et~al.} described a constraint based approach of rendering point cloud based data where the points are replaced by spheres or surface patches of approximate size~\cite{Leep11}. Another proxy based technique of rendering a dense 3-D point cloud data was proposed in~\cite{Sree12}, where the surface normal is estimated locally from the point cloud. Most of the methods are unable to handle scale changes during rendering and variable density of point cloud.

In order to enrich the experience of virtual world, many authors have incorporated surface properties like texture and friction in the haptic domain. Adi~ \emph{et~al.}~\cite{Adi09} introduced the technique of rendering the tactile cues from visual information using wavelet transforms. It was more realistic than primitive haptic texture rendering methods implemented using sine waves~\cite{Choi02} and Fourier series~\cite{Wall04}. But this technique was found to be less stable than the existing methods. Some authors, as in~\cite{Roma12}, have proposed data driven approaches to realistic haptic texture rendering. However, these methods are computationally expensive for real-time applications. Similarly, Richard~\emph{et~al.} presented a friction rendering model in haptics using modified Karnopp model~\cite{Rich02}. Hayward~\emph{et~al.}~\cite{Hayw00} developed a discrete implementation of friction exhibiting four friction regions: sticking, creeping, oscillating, and sliding. Harwin \emph{et~al.}~\cite{Harw04} proposed the friction cone algorithm for providing friction in haptic environments which also has a few shortcomings as suggested in~\cite{Meld04}. Hence, still much needs to be explored in this domain in order to augment the user's virtual experience.

Research in ecological acoustics imply that auditory feedback can effectively convey information about a number of object attributes such as its shape, size and material~\cite{Gave93}. The effectiveness of auditory sensations was studied by Lederman~\emph{et~al.}~\cite{Lede02} who have showed that sound plays a dominant role when a probe is used to interact with a surface as compared to the  case of direct contact with the bare fingers. We make use of this fact in order to improve the virtual immersion of the user interacting with the virtual environment using a haptic probe. Simultaneous audio rendering is required in our application where we try to render ancient cultural monuments where the supporting pillars have very interesting acoustic (read musical) properties.
\section{Proposed Proxy Updation Algorithm}\label{meth_sec}
The proposed system incorporates a proxy based algorithm to render a point cloud data. It is to be noted that the proxy is a point and not a sphere as it is popular in point cloud rendering techniques, thus allowing us to perceive surface textures. Let us first assume that one has the depth buffer data corresponding to 3-D models available. Issues related to transformation of the haptic space and scaling will be discussed in section \ref{render}.
 
In order to render the object in the haptic domain, we need to find the collision of HIP with the bounding surface and hence the penetration depth of HIP into the surface. The proposed algorithm tries to translate the proxy over the object surface in short steps during the haptic interaction such that each time it finds the most suitable proxy position which would provide the minimum distance between HIP and proxy and simultaneously applies the reaction force normal to the surface at the point of contact.
\begin{figure}[h]\label{tangent_fig}
\begin{minipage}{0.45\linewidth}
\centering
\includegraphics[width=0.99\linewidth]{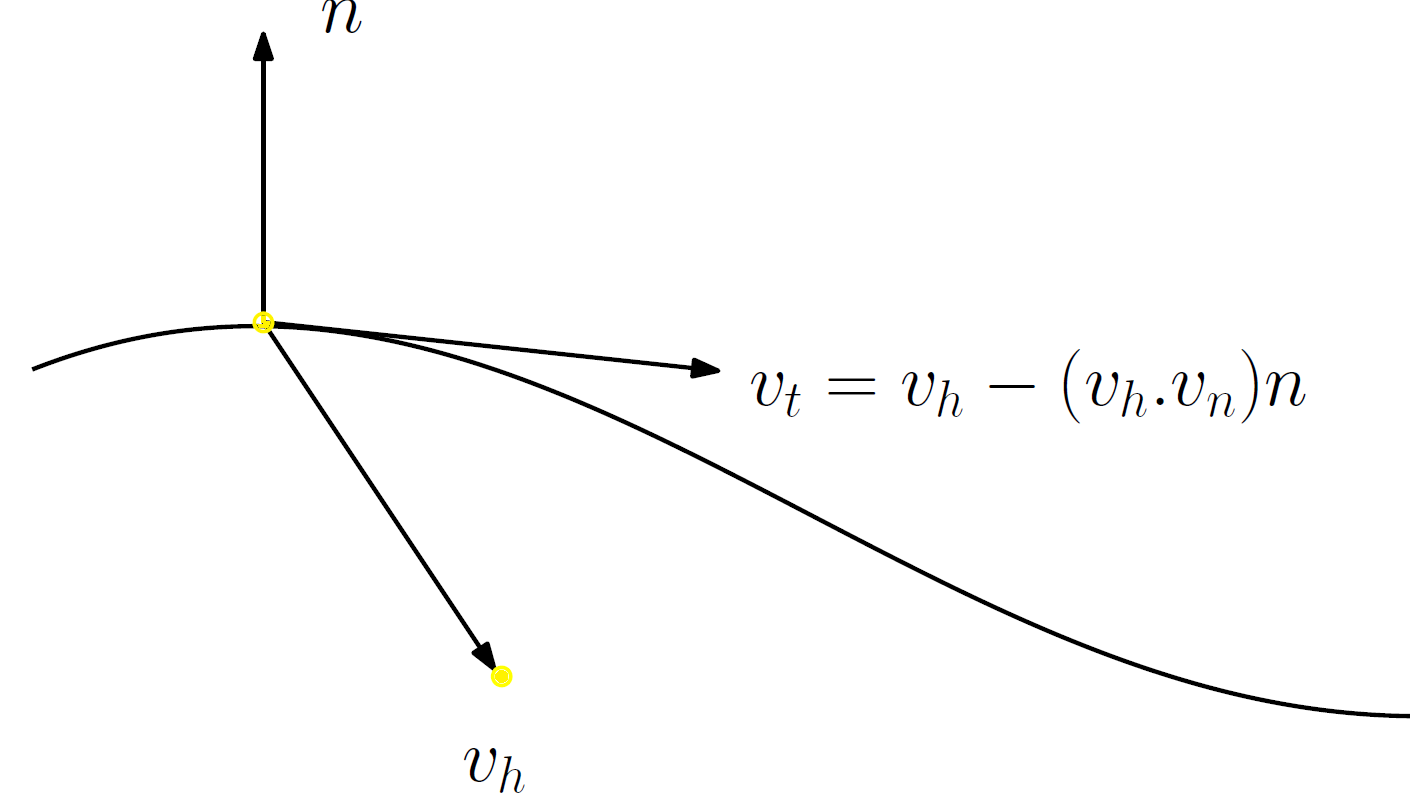}\\
\end{minipage}                     
\begin{minipage}{0.45\linewidth}
\centering
\includegraphics[width=0.99\linewidth]{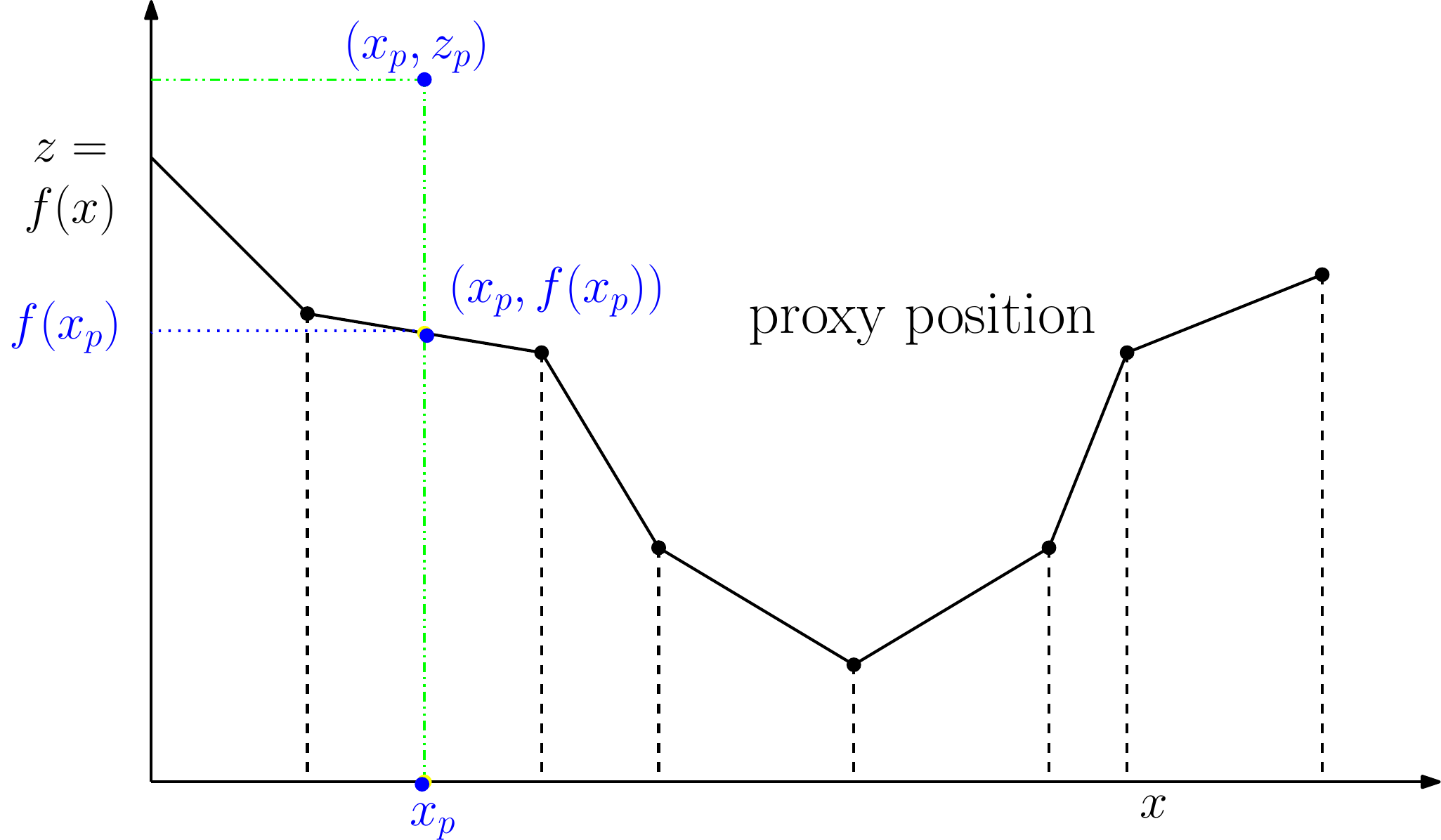}\\
\end{minipage}
\caption{(a) Illustration of tangent vector evaluation using the current proxy and HIP positions. (b) Illustration of surface approximation from depth values.}
\end{figure}

Fig.~\ref{tangent_fig}(a) illustrates the proxy movement during collision with an arbitrary object represented by a Monge surface. The Monge surface corresponding to a 3-D object is shown as a curve in $1-D$. HIP is shown with a yellow circle penetrated inside the object and the proxy is also shown with a yellow dot constrained to the surface. The vector  $\mathbf{v}_n$  represents the normal at the initial proxy position with $\mathbf{n}=\frac{\mathbf{v}_n}{|\mathbf{v}_n|}$ representing the unit normal and  $\mathbf{v}_{h}$  is the vector from proxy to the HIP and is given by  $\mathbf{v}_{h} = \mathbf{X}_{h}-\mathbf{X}_{p}$, where $\mathbf{X}_{h}$ and $\mathbf{X}_{p}$ represent the position vectors of HIP and proxy, respectively. The tangential vector in the plane of $\mathbf{v}_{n}$ and $\mathbf{v}_{h}$ is evaluated which provides a fast approximation of the direction for the proxy to move so that the distance between proxy and HIP can be minimized. When the proxy is moved continuously along the tangential direction on the curve, proxy will finally come to rest at a point where the angle between $\mathbf{v}_{n}$ and $\mathbf{v}_{h}$ is 180 degrees. The tangent vector $\mathbf{v}_t$ can be computed from $\mathbf{n}$ and $\mathbf{v}_{h}$ using equation \ref{eq3}.
\begin{equation}\label{eq3}
\mathbf{v}_t=\mathbf{v}_h-(\mathbf{v}_n.\mathbf{v}_h)\mathbf{n}.
\end{equation}
We use the following proxy update equation to translate the proxy along the tangent plane.
\begin{equation}\label{eq4}
\mathbf{X}_p^{(k+1)} = \mathbf{X}_p^{(k)}+ \rho\mathbf{v}_t^{k}.
\end{equation}
The parameter $\rho < 1$ and is arbitrarily chosen. As the value of $\rho$ increases, the proxy quickly converges, but it does not move close to the surface during the convergence. On the other hand if $\rho$ is very small, the proxy point moves close to the object surface, but needs a little more time to converge. As a matter of fact, the value of $\rho$ relates to the frictional force on the surface and should depend on the material property as explained in section 3.2. A small value of $\rho$ signifies larger surface friction. Hence the proposed method provides an easy way of including dynamic friction during rendering.

Once the proxy moves in the direction of tangential vector, it may deviate from the the boundary of the object. This deviation of the proxy from the boundary is avoided by projecting the proxy along $\mathbf{n}$ onto the surface before updating its position along the tangential direction. The proxy position update is performed within 1 ms of time, so that the user's interaction with the object through the haptic device is unhindered and is carried out at 1 kHz. Since the updated proxy location need not be on the chosen lattice for depth representation, the surface needs to be locally interpolated.
In case of $2-D$ depth data, we project the proxy onto the X-Y plane and the corresponding depth value is obtained by interpolating the neighbourhood depth values to form a continuous function $z=f(x,y)$ as illustrated in Fig.~\ref{tangent_fig}(b) for a $1-D$ function $z=f(x)$. Since the available points are sampled quite densely, bilinear interpolation is sufficient to find the bounding surface as shown in Fig.~\ref{tangent_fig}(b). In order to check the collision of HIP with the surface, we compare $z_p$ with the depth at the projected point $z=f(x,y)$ for a given proxy position $(x_p,z_p)$.  If $f(x_p)>z_p$, the proxy has touched the surface, otherwise it is free to move towards the HIP.
\begin{figure}[h]
\centering
\includegraphics[width = 0.75\textwidth]{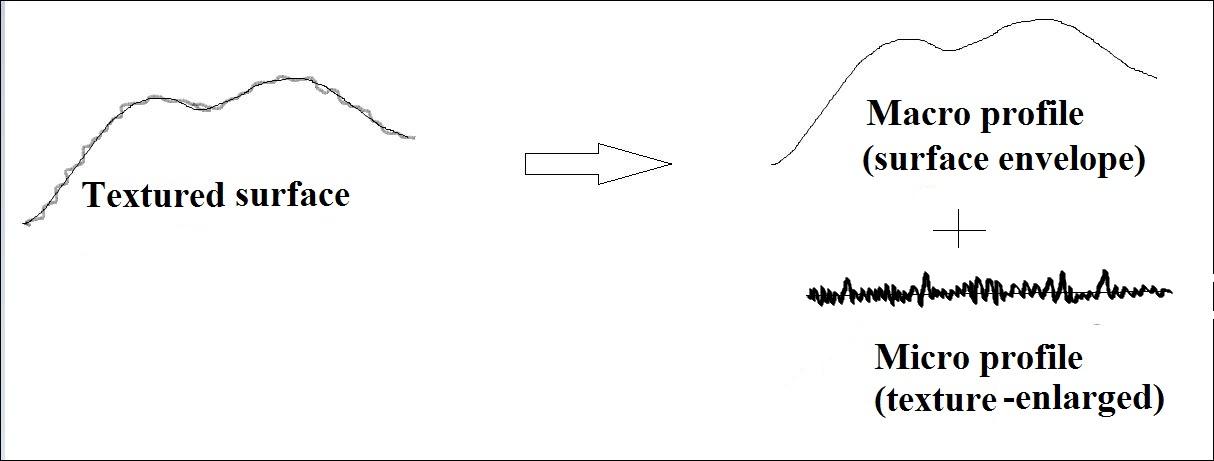}
\caption{Envelope subtraction from a textured surface.}
\end{figure}
\label{envelope}
\subsection{Recovery of surface texture}
Any real or virtual surface can be represented as a sum of a general shape or envelope of the surface and the minute surface variations called texture. The latter provides a realistic feel to an otherwise slippery envelope. Hence the texture component of the surface can be extracted by subtracting the envelope (low frequency component) from the surface as illustrated in Fig.~\ref{envelope}. The figure shows how a  textured Monge surface can be represented as a combination of a macro profile (general geometry or shape) and micro profile (texture). In this work, a bilateral filter is used for the purpose of envelope subtraction. Bilateral filtering~\cite{Toma98} provides simple and non-iterative edge-preserving data smoothing. The bilateral filter takes a weighted sum of the depth map at a local neighbourhood at the lattice; the weights depend on both the spatial distance and similarity in depth values. In this way, edges are preserved well while ``noise'' is averaged out. The bilateral filtered output $f_b(x,y)$ of the depth data $f(x,y)$ is obtained from pixels $(\tilde{x},\tilde{y})$ in the neighbourhood as shown in the following equation.
\begin{figure}[h]\label{bilateral_fig}
\begin{minipage}{0.45\linewidth}
\centering
\includegraphics[width=0.6\linewidth]{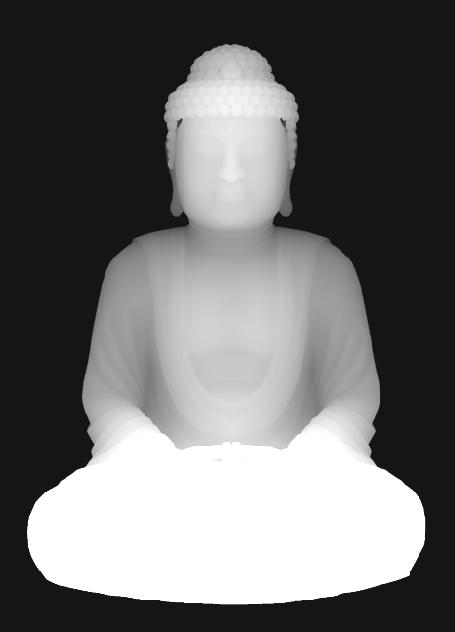}\\
\end{minipage}                     
\begin{minipage}{0.45\linewidth}
\centering
\includegraphics[width=0.6\linewidth]{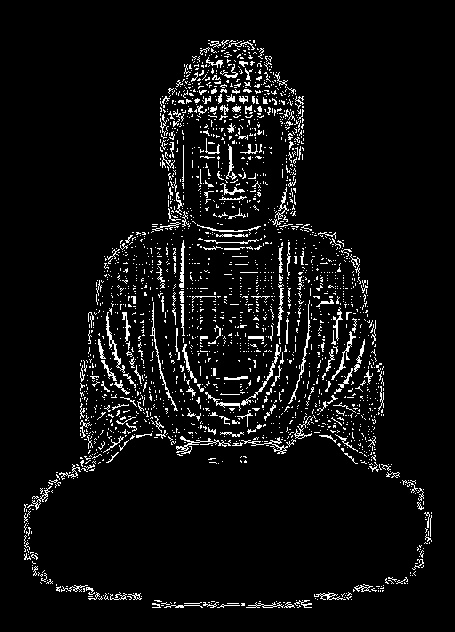}\\
\end{minipage}%
\caption{Illustration of texture retrieval from depth data using a bilateral filter:(a) Depth image corresponding to 3-D model of Buddha (b) Extracted texture details (Data Courtesy: \emph{www.archibaseplanet.com}).}
\end{figure}
\begin{equation}
f_b(x,y)=\frac{1}{W(x,y)}\sum_{\tilde{x}}\sum_{\tilde{y}}{G_{\sigma_S}(x-\tilde{x}, y-\tilde{y})G_{\sigma_R}(f(x,y)-f(\tilde{x},\tilde{y}))f(\tilde{x},\tilde{y})}
\end{equation}
where $G(x,y)$ is a Gaussian kernel and $\sigma_R$ and $\sigma_S$ represent the spread in amplitude values and spatial distances, respectively. The term $W(x,y)$ is a normalization factor. Although this is a non-linear filter, computationally efficient algorithms exist to obtain the filtered output~\cite{Toma98}. The bilateral filtered output $f_b(x,y)$ of the 3-D object is subtracted from the depth data $f(x,y)$ so as to obtain the texture component alone. Hence 
\begin{equation}
h(x,y)=f(x,y)-f_b(x,y).
\end{equation}
Fig.~\ref{bilateral_fig}(a) shows the depth image of a 3-D model and Fig.~\ref{bilateral_fig}(b) shows the extracted texture from the depth image using a bilateral filter. As explained earlier, bilateral filter provides edge-preserving smoothing and hence its output is the smoothed depth data without texture details.
\subsection{Incorporation of surface friction}\label{friction}
\begin{figure}
\centering
\includegraphics[width=0.5\textwidth]{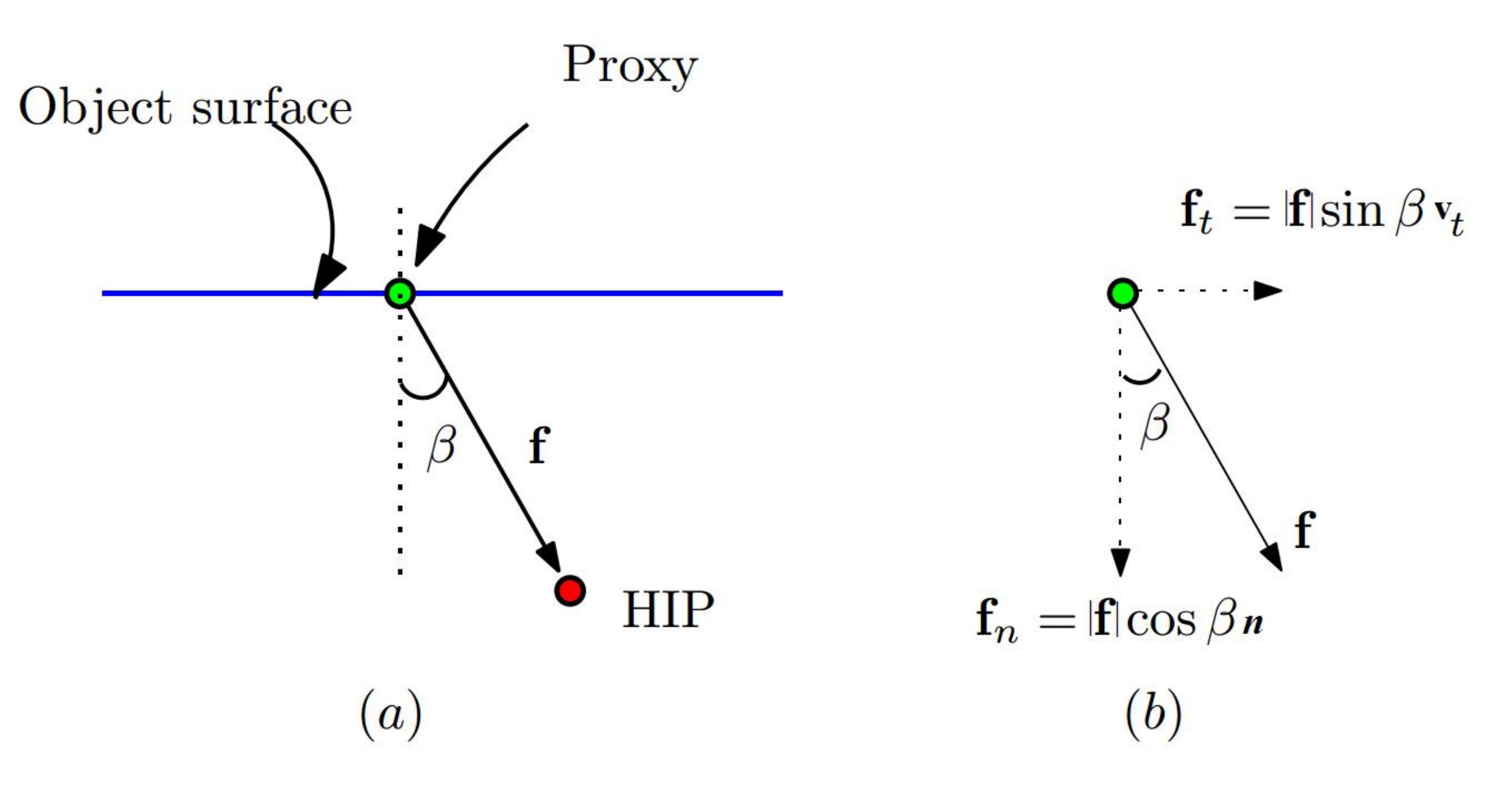}
\centering
\caption{Illustration of calculation of the resultant vector for the proxy movement while incorporating surface friction.}
\label{friction_fig}
\end{figure}
Haptic rendering techniques that do not consider the surface friction induce the feeling of a very smooth and slippery surface which is not the case in practice. Adding friction along with the texture details to the model provides a more realistic feeling of the surface. Let $\mathbf{f}$ be the reaction force on the haptic device as shown in Fig.~\ref{friction_fig}. The component of the applied force normal to the surface $\mathbf{f}_n$ is given by $|\mathbf{f}|\cos\beta$. Similarly the tangential force on the surface is $|\mathbf{f}|\sin\beta$. The magnitude of the retarding
force on the proxy is proportional to the normal force $\mathbf{f}_n$. If $\mu_s$ denotes the static friction coefficient, proxy is in static contact with the surface as long as $|\mathbf{f}_t| < \mu_s |\mathbf{f}_n|$. During haptic interaction, dynamic friction exerts a retarding force on the proxy while moving on the surface of the object. $\mu_d$ which denotes the coefficient of dynamic friction is made proportional to the resultant curvature of texture component at each point of the rendered surface since friction depends on the surface property of the material. Since the curvature encodes the unevenness of a surface very well, we use $\mu_d$ proportional to the curvature. Now, the magnitude of the retarding force is given by $\mu_d|\mathbf{f}_n|$ and is in a direction opposite to $\mathbf{f}_t$.  The curvature is computed as the resultant of mean and Gaussian curvatures whose magnitudes are given by the following equations. The physical significance of mean curvature($H$) is the first variation of the surface area and Gaussian curvature($K$) represents the local convexity.  Hence the use of resultant curvature as a representative of $\mu_d$ can be justified as it represents the amount of ``bending" at each point on the surface~\cite{math00}.
\begin{equation}
H = \frac{{h}_{xx}(1+{h}_y^{2})+ {h}_{yy}(1+{h}_x^{2})-2{h}_{xy}{h}_{x}{h}_{y}}{2(1+{h}_x^{2}+{h}_y^{2})^\frac{3}{2}}
\end{equation}
\begin{equation}
K=\frac{{h}_{xx}{h}_{yy}-{h}_{xy}^{2}}{(1+{h}_x^{2}+{h}_y^{2})^{2}}
\end{equation}
where the parameters $h_{x}$, $h_{y}$ are first partial derivatives of the texture component $h(x,y)$ of the surface \textit{w.r.t} $x$ and $y$ axis. Similarly $h_{xx}$, $h_{yy}$ are second partial derivatives of  $h(x,y)$ \textit{w.r.t} $x$ and $y$. The dynamic friction coefficient $\mu_d$ is computed as follows:
\begin{equation}
\mu_d=\frac{1}{R\sqrt{H^{2}+K^{2}}}
\end{equation} 
where $R$ is the radius of the largest inscribable sphere within the haptic work space. Since $R$ is much larger than the radius of curvature in the micro texture, usually $\mu_d<1$.
The resultant force $\mathbf{f}_r$ on the proxy is given by $\mathbf{f}_t - \mu_d \mathbf{f}_n$.
\begin{eqnarray}\label{math_eq3}
\mathbf{f}_r &=& \mathbf{f}_t(1-\mu_d\cot\beta) \hspace{0.4cm} if \hspace{0.8cm}|\mathbf{f}_t| \geq \mu_s |\mathbf{f}_n|\nonumber\\ 
                     &=& 0 \hspace{0.4cm} otherwise 
\end{eqnarray}

Equation \ref{math_eq3} provides a direct description of the frictional force. For proxy based haptic rendering, such an equation can easily be incorporated  while defining the proxy movement. Comparing equations (\ref{math_eq3}) with (\ref{eq4}), we observe that $\mathbf{f}_r$ is proportional to $\mathbf{v}_t$ . Hence the parameter $\rho$ with friction is given by $\rho_f = \rho(1-\mu_d\cot\beta)$. Then the proxy update equation is given by:\\
\begin{equation}\label{eq5}
\mathbf{X}_p^{(k+1)} = \mathbf{X}_p^{(k)}+ \rho(\mathbf{1-} \mu_d\cot\beta^{k})\mathbf{v}_t^{(k)}
\end{equation}
where $k$ denotes the iteration number. 
\section{Multimodal Rendering}\label{render}
Haptic rendering involves generating force feedback in order to provide the sensation of touch to the users. Any haptic rendering algorithm would include the following two steps:
{
\begin{enumerate}
\item detection of collision of the HIP with the object.
\item computation of force feedback if a collision is detected. 
\end{enumerate}
}
If $z_p<f(x_p,y_p)$ in Fig.~\ref{tangent_fig}(b) then the proxy has touched the object and a force needs to be fed back to the user through the haptic device.
Subsequently as explained in section~\ref{lit_sec}, the reaction force is computed as $\mathbf{F}=-k\mathbf{x}$ where $k$ is the Hooke's constant, and $\mathbf{x}$ is the current penetration depth given by $\mathbf{x}=\mathbf{X}_h-\mathbf{X}_p$, where $\mathbf{X}_h$ is the HIP position and $\mathbf{X}_p$ is the proxy position. 

For a combined hapto-visual rendering, the 3-D object surface is displayed as a simple quad mesh formed out of the depth values in OpenGL. We have opted for the mesh-based graphical display in order to give a better perception to the viewer since using point cloud data for graphic display would result in gaps in the visually rendered image. We have used the stereoscopic display technique for creating the effect of depth in the image by presenting two offset images  on the screen corresponding to two different points of projection. This provides a 3-D perception to the user. Anaglyphic glasses can be used to view such displays.
\subsection{Rendering at Different Scales and Orientations}
\begin{figure}[h]
\centering
\begin{subfigure}[]{\includegraphics[width=.4\linewidth] {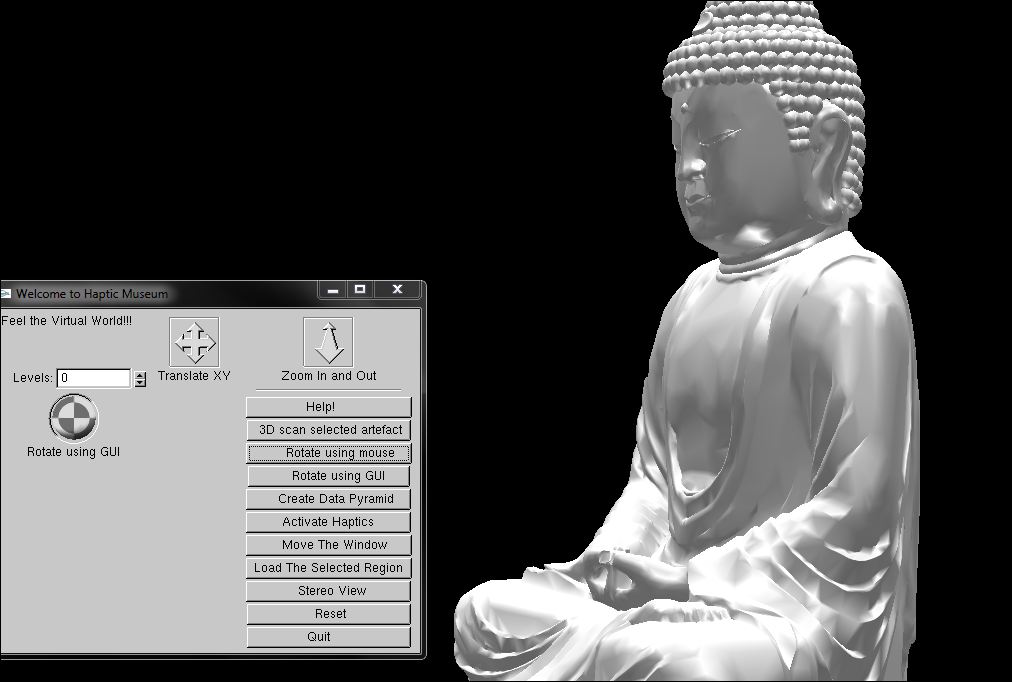}
   \label{drag_fig1}
 }
\end{subfigure}\hfill
 \begin{subfigure}[]{\includegraphics[width=.4\linewidth]{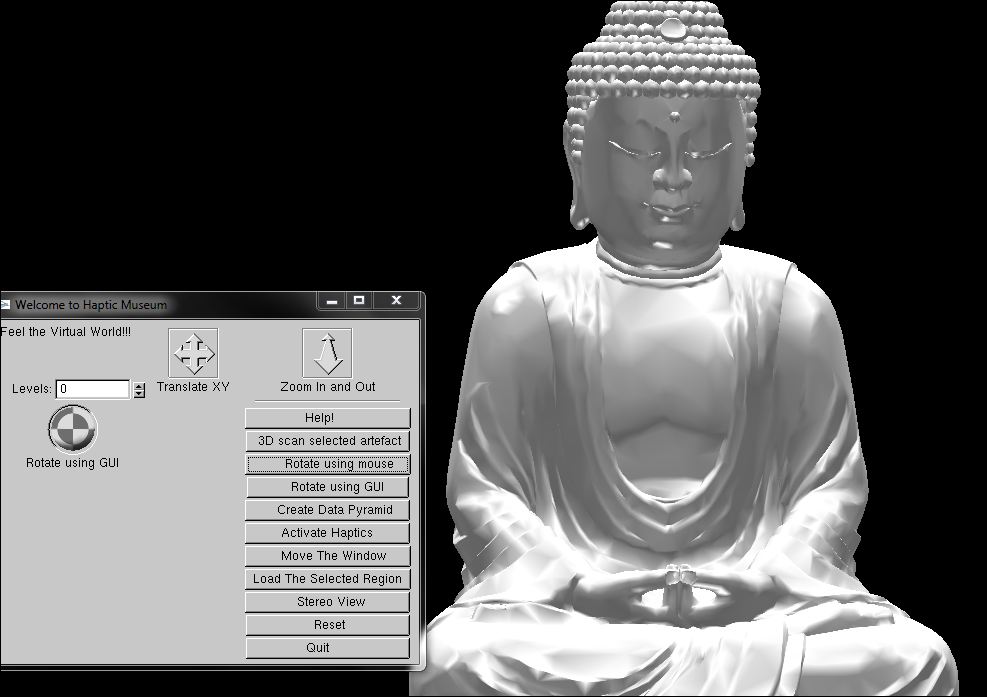}
   \label{drag_fig2}
 }
\end{subfigure}\hfill
 \begin{subfigure}[]{\includegraphics[width=.4\linewidth]{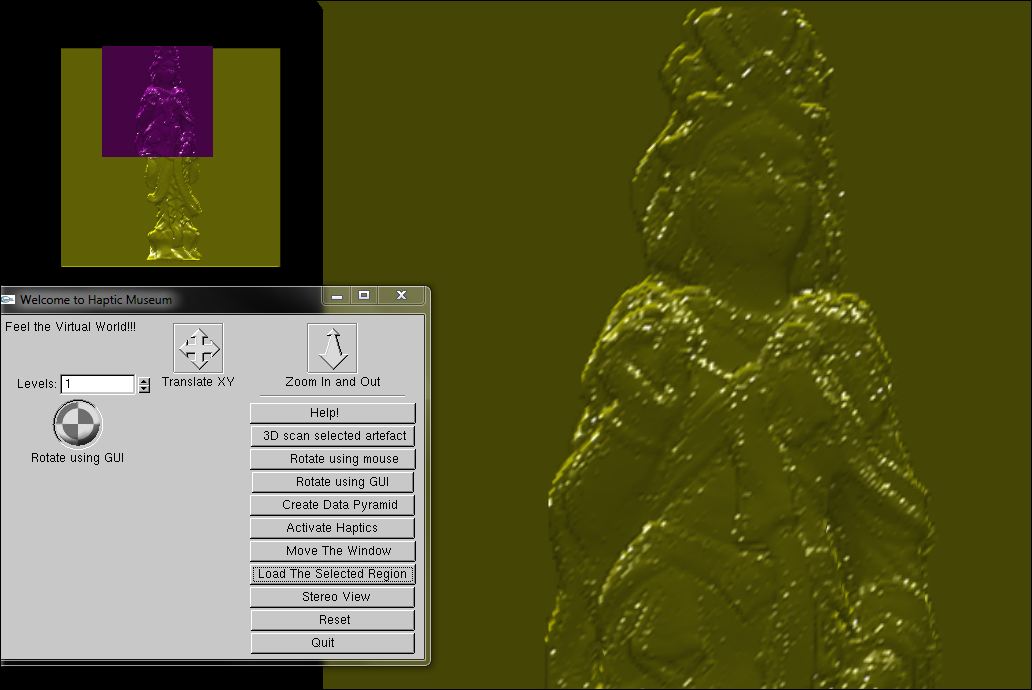}
   \label{drag_fig3}
 }
\end{subfigure}
\caption{Illustration of user defined selection of orientation (before depth scan): 3-D model of Buddha (a) before rotation (b) after rotation. (c) Illustration of user defined selection of level of details (at run-time) for another model(Data Courtesy:\emph{www.archibaseplanet.com}).}
\label{drag_fig}
\end{figure}
In practice, 3-D objects come at various physical scales and orientations. In a virtual environment, one should be able to experience objects of all sizes at different scales to get a sense of overall structure to finer details from the same data set. Hence, we have implemented adaptive scaling in both graphical and haptic domains. In order to scale the surface we resize depth data of resolution $N \times N$ depending on the level user selects, with $N \times N$ as the finest level. If we load the level $N \times N$ into the haptic space the full object can be rendered visually as well as haptically. Users can select the level as well as the region of interest at run time either using buttons in the haptic device or using keyboard functions. Additionally, we have developed a graphical user interface for easy acessibility. Fig.~\ref{drag_fig} illustrates the selection of scale and orientation by the user. Fig.~\ref{drag_fig}(b) shows the rotated version of Fig.~\ref{drag_fig}(a). Fig.~\ref{drag_fig}(c) illustrates how user can select different levels of details and the corresponding textural component is shown.

The user can select the desired orientation before the depth scan using either the mouse controls or using the graphical user interface. Further, depending on the scale selected by the user, the corresponding depth data is dynamically loaded into the active haptic space and an appropriate haptic force is rendered. As only a limited subset of data is loaded, the rendering is very fast. In general, at higher levels of resolution, the user should be able to view finer details. The haptic force also varies accordingly. Hence in order to incorporate realistic haptic and graphic perception, we need to appropriately scale the depth values at each level of depth map. Further, trying to map a large physical dimension over a small haptic work space (typically about 4 inch cube of active space) leads to a lot of unwanted vibrations (something similar in concept to aliasing) during rendering. Hence the depth values need to be smoothed before being down-sampled and mapped into the haptic work space. Multi-scale data generation for procuring different levels of details is performed based on ~\cite{Gluc06}.
\subsection{Audio Rendering}\label{hampi}
\begin{figure}[h]
\begin{minipage}{0.45\linewidth}
\centering
\includegraphics[width=0.9\linewidth]{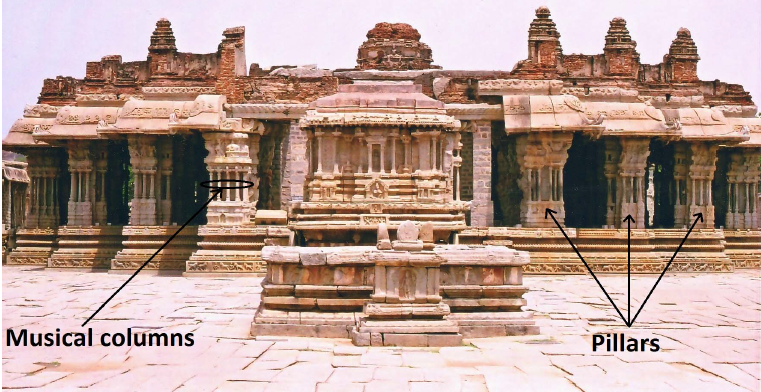}\
\end{minipage}
\begin{minipage}{0.45\linewidth}
\centering
\includegraphics[width=0.9\linewidth]{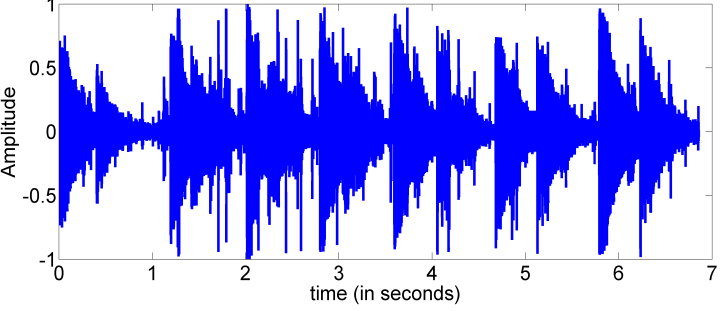}\\\end{minipage}
\caption{(a) Front view of Vitthala temple, Hampi(courtesy:~\cite{Pati12}). (b) Plot of an audio file recorded from the musical pillar. (courtesy:~\cite{Pati12}).}
\end{figure}
Sounds are incorporated in the rendering framework by playing appropriate audio files based on the position of haptic probe inside the virtual environment. A demonstration of this technique is given in the attached video using the 3-D model of musical pillars at Vitthala temple, Hampi which is an early 14\textsuperscript{th} century world heritage site. The pillars in this temple have musical columns which produces distinct sounds when struck. The temple consists of 56 pillars which are monolithic sculptures each having granite stone columns of height 10 feet as shown in Fig.~\ref{hampi}(a). Each major pillar is surrounded by 7 minor pillars that can reverberate at 7 primary notes of Indian classical music. We have used the recorded audio files of these pillars (data courtesy: \emph{www.daiict.com}) in our work. A sample audio file is shown in Fig.~\ref{hampi}(b). A single wavelength of the sound file was extracted from the above audio file and was played back whenever the haptic probe touched the virtual pillar. During audio rendering, one could synthesize various types of sounds synthetically. However, we provide the real data from the actual heritage site to provide a real feel of the musical pillars. During haptic rendering whenever a collision takes place for the first time with a specific pillar, the corresponding note is played, the volume of which is made proportional to the rendered force.   
\section{Results}\label{results}
The proposed method was implemented in visual C++ in a Windows XP platform with an Intel i5 CPU @ 2.66 GHZ with 2 GB RAM. For obtaining texture details, the depth data obtained from OpenGL depthbuffer is fed to a bilateral filter using OpenCV inbuilt functions and the output from the filter is subtracted from the original depth image. We have experimented with various models of 3-D objects and a few of them are displayed below. Fig.~\ref{rack_fig} shows the model of Ganesh, visually rendered in OpenGL. For haptic rendering we use HAPI library. The blue ball represents the position of the proxy constrained to lie on the surface. The discrete position in the model is displayed in a fixed $200\times200$ haptic space. The size and spatial resolution of the model depend on two factors: the active space of the haptic device used to render the model, and the resolution at which the model should be displayed. We use a 3-DOF haptic device, NOVINT FALCON with a 4 inch cube of active space. While interacting with the object haptically, the average proxy update time was found to be 0.0056 ms which is much faster than the required upper bound of 1 ms, and hence the user has a very smooth haptic experience. The average time required for dynamic data generation and loading it into the haptic space depends on the resolution of input depth data and it was observed to be 
around 0.5 s and 0.02 s, respectively, for depth data with a resolution of $800\times800$. We also carry out the rendering at finer levels of details by successively zooming into the object.
\begin{figure}[h]
\centering
\begin{subfigure}[]{\includegraphics[width=.45\linewidth] {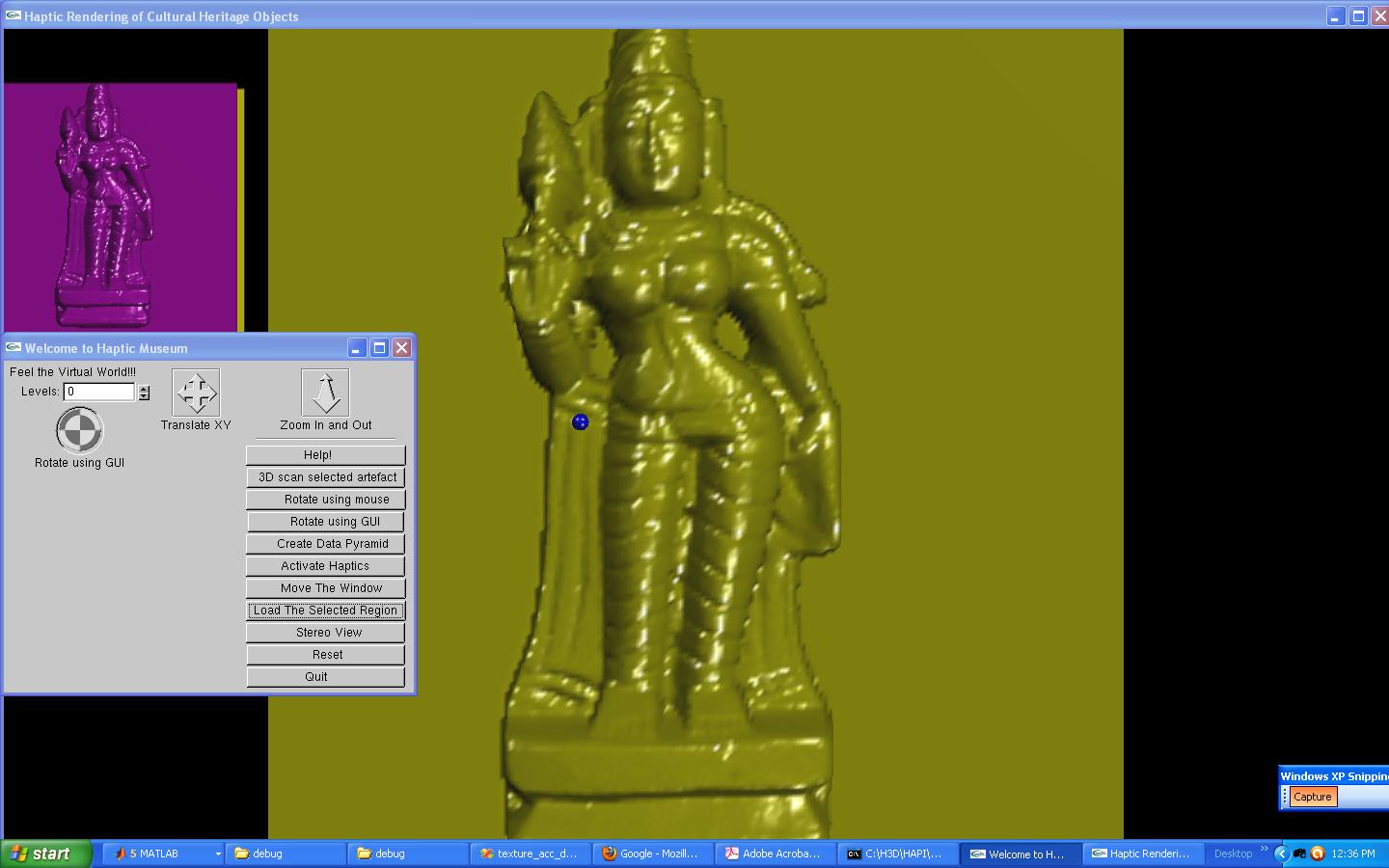}
   \label{rack_fig1}
 }
\end{subfigure}\hfill
 \begin{subfigure}[]{\includegraphics[width=.45\linewidth]{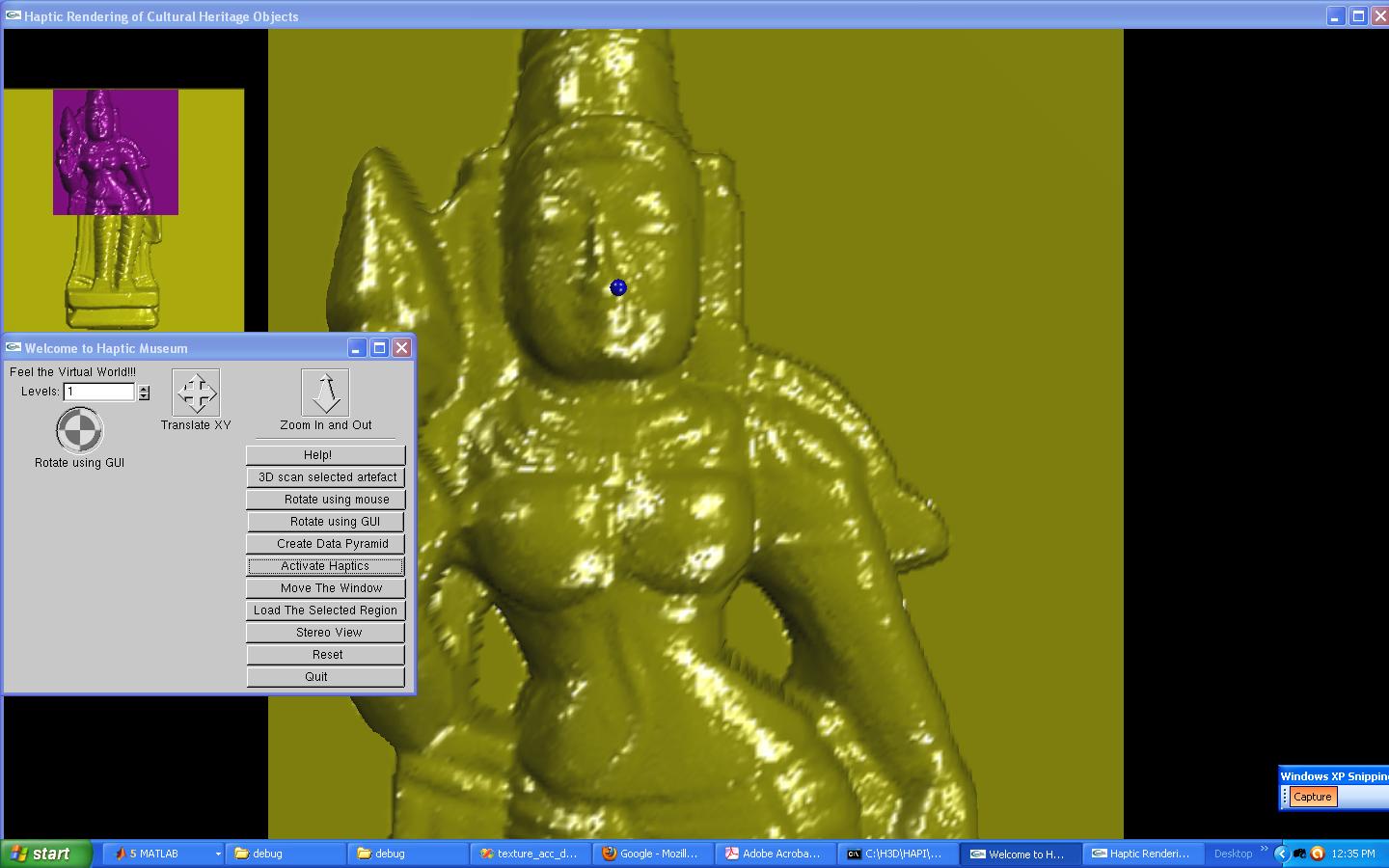}
   \label{rack_fig2}
 }
\end{subfigure}\hfill
 \begin{subfigure}[]{\includegraphics[width=.45\linewidth]{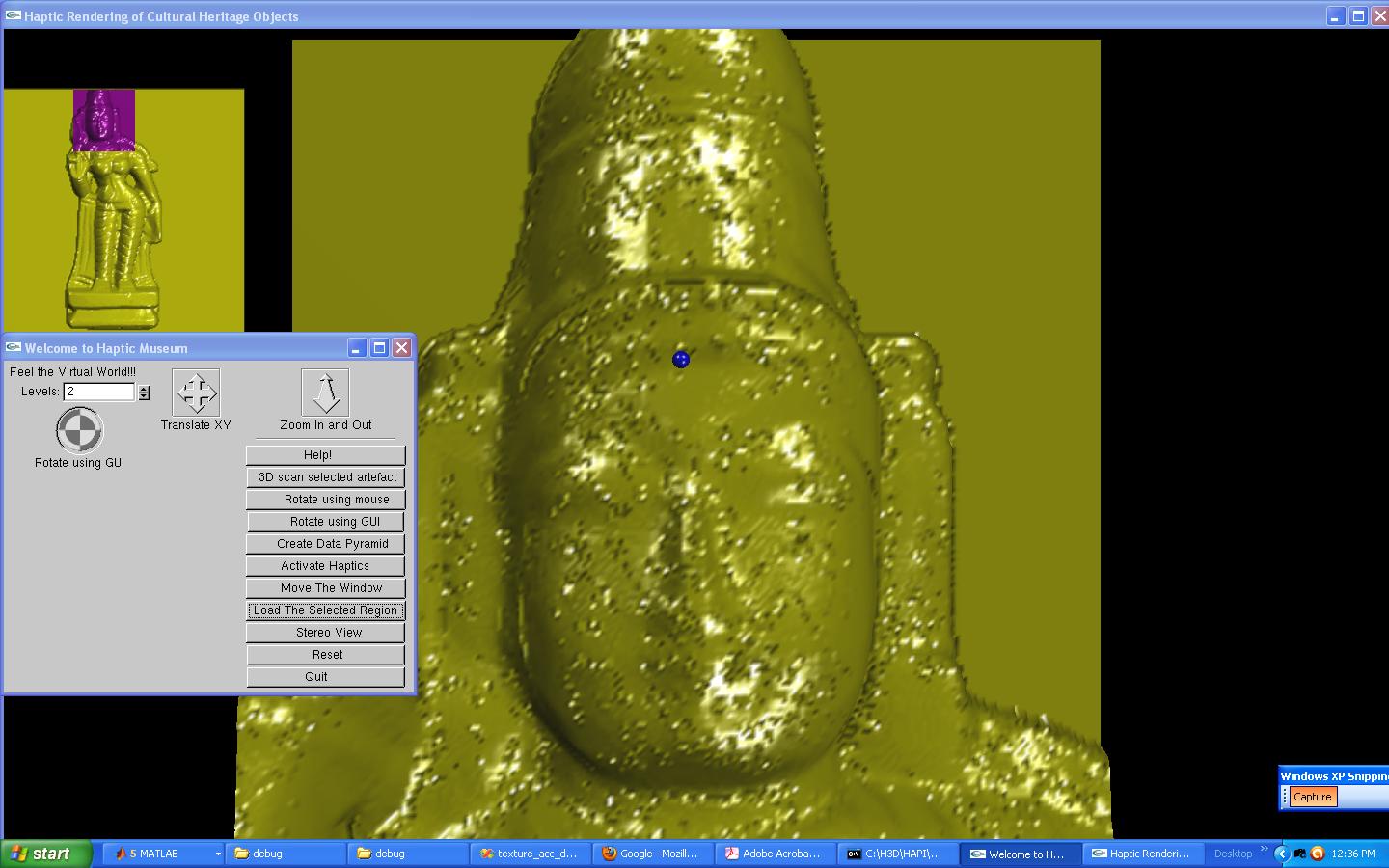}
   \label{rack_fig3}
 }
\end{subfigure}
\caption{3-D model of a statue with texture: at (a) lowest level of details (b) at double the resolution and (c) at the finest resolution. (Data Courtesy: \emph{www.archibaseplanet.com}).}
\label{rack_fig}
\end{figure}
In above cases, each figure consists of two parts where the left part is the reference for 
the users to select the part of the object they wish to explore haptically as shown in Fig.~\ref{drag_fig} and Fig.~\ref{rack_fig}. The right part of 
the figure corresponds to the selected region at the appropriate resolution for haptic rendering.
Fig.~\ref{rack_fig}(b) shows the scaled up version of Fig.~\ref{rack_fig}(a). It is quite clear from 
Fig.~\ref{rack_fig}(c) that the users are able to feel even minute details of the sculpture and 
have visual perception of closeness in depth. Hence they can have a more realistic experience. Furthermore, the audio-playback feature with respect to the musical pillars augmented the user experience. However, we observe that there is a small time lag between the haptic and audio rendering due to data access time to open the stored audio files. For a small number of musical notes, these audio clips can be stored in RAM to circumvent this problem.
\subsection{Validation of proposed method}
\begin{figure}
  \centering
  \includegraphics[width=0.5\textwidth]{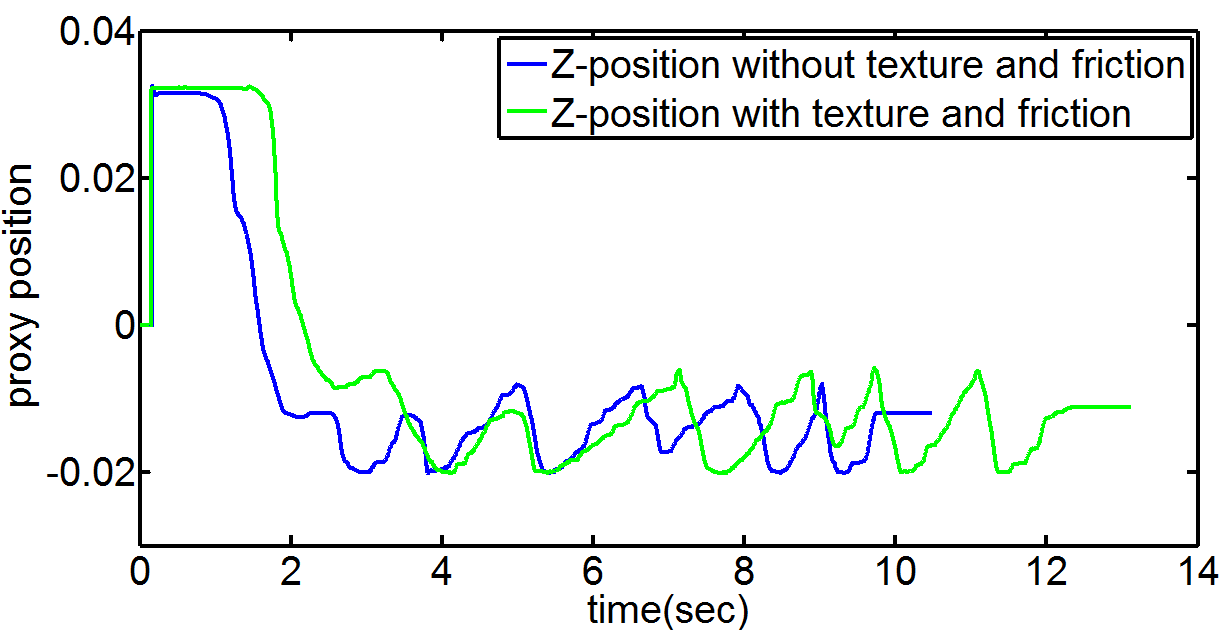}	
	\centering
	\caption{Plot of $z$-position of proxy as a function of time to illustrate the effect of surface friction.}
	\label{proxy_1_fig}
\end{figure}
Validation of result is not quite easy during haptic rendering. Authors in~\cite{Sreeni12} propose a validation technique using a standard 3-D sphere model. But, since rendering of friction in our work involves computation of resultant curvature, the use of a spherical model (having a fixed curvature) is not justified.  We demonstrate a validation technique using a known sinusoidal surface. Fig.~\ref{proxy_1_fig} shows the $z$-component of proxy as a function of time while haptically interacting with the depth data. The green line and the blue line in the figure show the $z$-component of the proxy point with and without texture and friction, respectively. We observe that the proxy position converges, but with a time-lag in the former case, as expected due to surface effects of texture and friction. The shapes of both the curves are quite similar except for the time lag.
\begin{figure}[h]
\centering
\begin{subfigure}[]{\includegraphics[width=.5\linewidth] {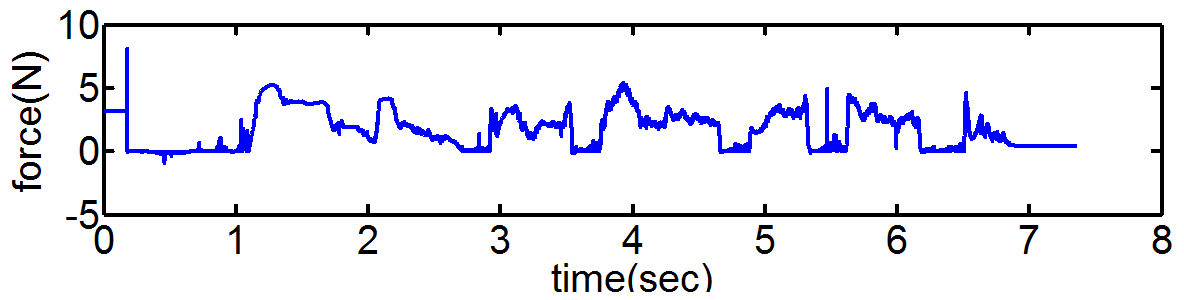}\label{force_fig1}}
\end{subfigure}\hfill
\begin{subfigure}[]{\includegraphics[width=.5\linewidth]{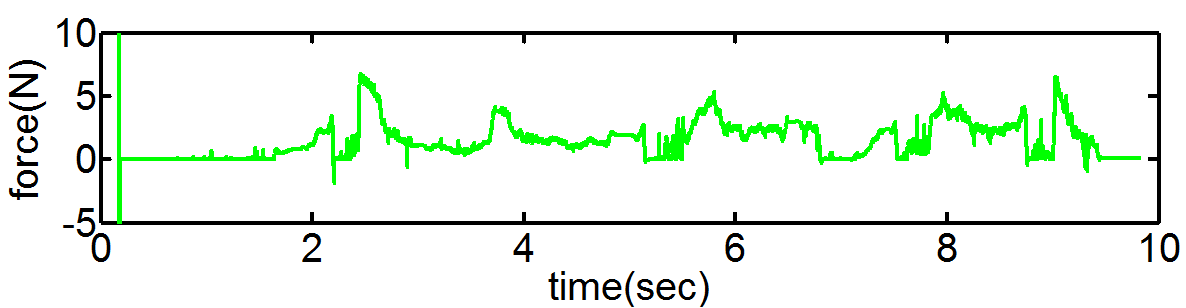}\label{force_fig2}}
\end{subfigure}
\caption{Force-vs-time graph for a particular interaction with the depth data (a) without surface effects and (b) with surface effects.}\label{force_fig}
\end{figure}
Fig.~\ref{force_fig}(a) shows the reaction force versus time during haptic interaction with a known sinusoidal surface without incorporating the surface effects. In free space the HIP and proxy positions are almost the same and hence the reaction force on the haptic device is zero for initial few seconds as illustrated in figure. As the HIP penetrates the object, the proxy stays on the surface according to the iteration method discussed in section \ref{meth_sec}. The proxy point moves continuously during interaction, whenever there is a change in HIP position and appropriate reaction force is fed back to the haptic device. The force-time graph in Fig.~\ref{force_fig}(b) illustrates how the net reaction force fed back to the haptic device is effectively delayed by constraining the proxy movement which gives the perception of friction to the user.
\begin{figure}[h] 
\includegraphics[width=0.7\textwidth]{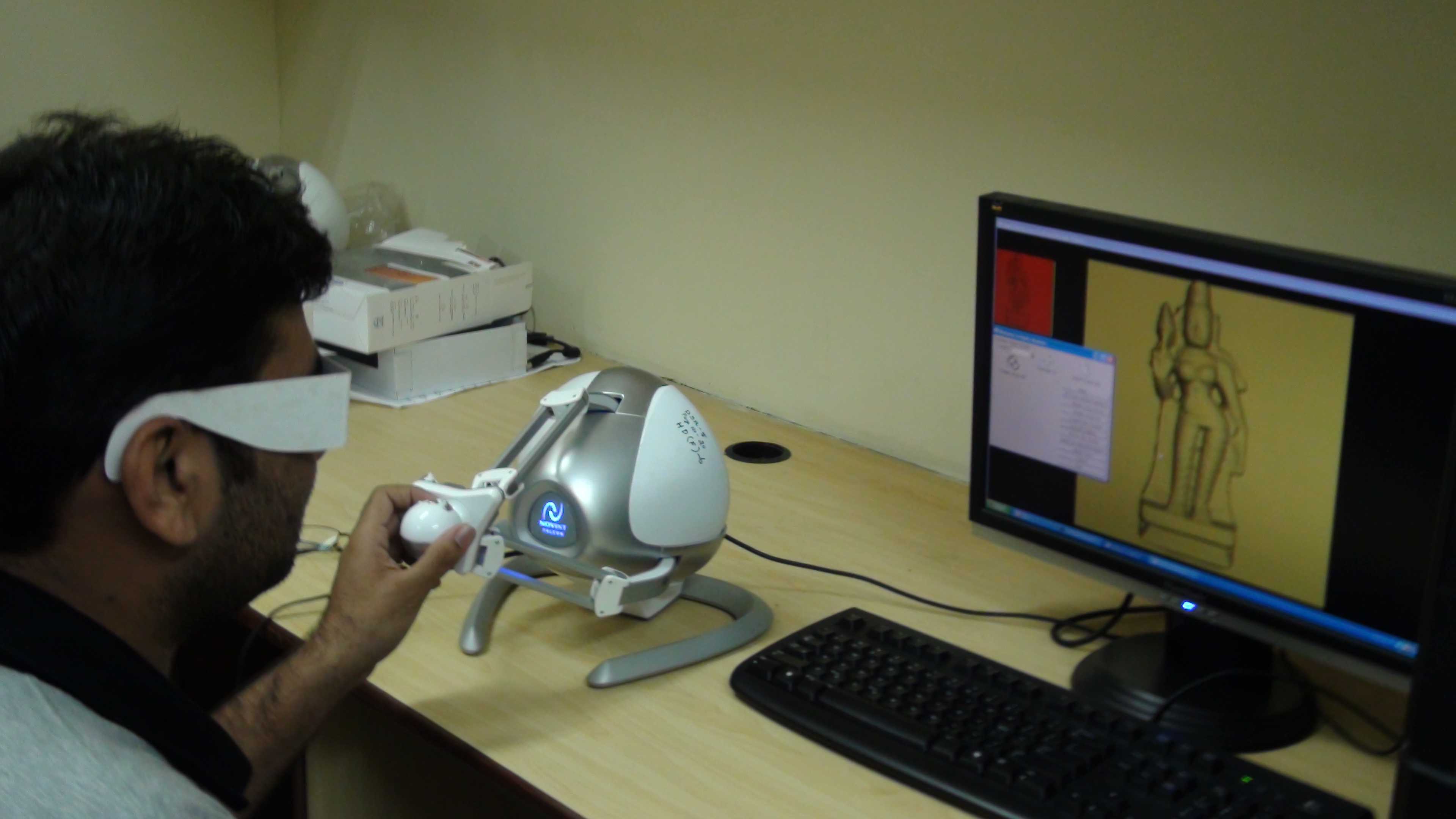}
\centering
\caption{Illustration of hapto-visual immersion of a subject for a 3-D 
object. On the left, the user wearing anaglyphic glasses is holding the FALCON haptic device 
while interacting with the virtual 3-D object displayed on the screen.}
\label{setup_fig}
\end{figure} 

In Fig.~\ref{setup_fig}, we show the actual set up of our rendering framework. A user wearing the anaglyphic glasses watches the stereoscopic visual rendering of an object and at the same time haptically interacts with the object through the Falcon device. This provides an excellent hapto-visual immersion of the subject into the virtual object. However, for the visually impaired users, the selection of scale and the location for rendering cannot be based on the small navigation window on the screen.  For such subjects, we use the buttons available on the haptic device for the user to explore the object at different scales and locations. We also conducted a user survey on the proposed virtual set-up in order to understand how realistically the users can perceive the shape and surface properties of the virtual models. All subjects were made to sit at a desk with the virtual environment displayed in front of them. Then they were asked to explore the virtual system by grasping the haptic device. After an exploration time of 1 minute, each user was asked to rate the realism of the virtual surface in terms points from 1-10 (with 10 being the highest rating). A total of 12 subjects (8 males and 4 females) volunteered for user study. All the users were in the age group of 18 to 40. The survey resulted in an average user rating of 7.23 (out of 10) for the realism of virtual environment. We propose to conduct richer user study in future including distribution and statistical meaning of user ratings.
\section{Conclusions}\label{conclusions}
In this work we have proposed a new technique for multi-modal rendering of 3-D objects represented as a dense depth map data. Rendering of surface properties like texture and friction is found to enrich the user's experience in the virtual world. We include scalability, rotation, translation and stereoscopic display of 3-D models as additional features to enhance the realism in experience. Presently, we have integrated audio rendering with respect to a single spatially segmented 3-D object i.e., the musical pillars at Hampi, India. In future, we propose to include continuous audio rendering in a more generalized framework by annotating the acoustic property at each point in the 3-D model. We conducted experiments with several 3-D models. We also conducted an user survey on a few subjects and observed that hapto-visual and auditory rendering of virtual 3-D models using the proposed method greatly augmented the user's experience.

\vspace{3mm}
\noindent {\bf Acknowledgement}. 
The authors would like to thank DST for the grant provided on the Indian Digital Heritage Project and MCIT for the grant on perception engineering. The authors would also like to thank Prof. Manjunath Joshi and his team for providing us with the audio signals of musical pillars ate Hampi which is the input to our proposed rendering framework.
%%===========================================================
\bibliographystyle{splncs}
%

%===========================================================

\end{document}